\documentclass[twocolumn]{aastex631}

\begin{document}

\title{The energy content of  Alfv\'en waves in the stratified solar atmosphere}

\correspondingauthor{Roberto Soler}
\email{roberto.soler@uib.es}

\author[0000-0001-6121-7375]{Roberto Soler}
\affiliation{Departament de F\'{i}sica, Universitat de les Illes Balears, E-07122, Palma de Mallorca, Spain}
\affiliation{Institute of Applied Computing \& Community Code, Universitat de les Illes Balears, E-07122, Palma de Mallorca, Spain}

\begin{abstract}

Alfv\'en waves propagating in a vertically stratified plasma, such as those travelling from the solar photosphere to the corona, are partially reflected due to the  gradient in the Alfv\'en speed. Wave reflection naturally results in the superposition of upward- and downward-propagating waves. A simple analytic model demonstrates that this superposition leads to the non-equipartition of kinetic and magnetic energies in the Alfv\'en wave perturbations and slows down the  net energy transport. A numerical model of Alfv\'en wave propagation in the lower solar atmosphere reveals significant wave reflection below the transition region, leading to highly variable kinetic and magnetic energy content in the lower chromosphere. At higher altitudes, the kinetic energy eventually dominates, depending on the wave frequency. The velocity at which net energy propagates upward is significantly smaller than the local Alfv\'en speed throughout the chromosphere. Consequently, the commonly used expression for unidirectional Alfv\'en waves in a uniform plasma, which relates the energy flux to the kinetic energy density, is not generally applicable in the stratified lower solar atmosphere and cannot be reliably used to estimate the energy content of observed waves. A generalized expression is given, incorporating correction factors that account for wave reflection and energy non-equipartition. The applicability of the expression is discussed.

\end{abstract}


\section{Introduction} \label{sec:intro}

High-resolution observations have clearly demonstrated the ubiquitous presence of Alfv\'enic waves in the solar atmosphere \citep[see the comprehensive review by][and references therein]{morton2023}. As originally proposed by \citet{alfven1947}, the waves are believed to be driven by the continuous plasma motions at the photosphere \citep[see, e.g.,][]{bonet2008,chitta2012,Stangalini2013} and propagate upward, transporting energy to the overlying atmosphere and solar wind. In this scenario, the chromosphere and transition region pose a formidable barrier for the upward-propagating Alfv\'en waves, which are expected to be strongly reflected due to the abrupt density gradient. The reflection process would naturally produce downward-propagating waves. Indeed, the presence of counter-propagating waves  in the chromosphere (a potential signature of reflection) has been reported in a number of observations \citep[e.g.,][]{okamoto2011,Liu2014,Jafarzadeh2017,Yoshida2019,bate2022,Murabito2024}. In addition to reflection, chromospheric Alfv\'enic waves are also affected by dissipation due to partial ionization effects \citep[see the review by][]{soler2024}. Despite the strong chromospheric filtering associated with both reflection and dissipation, it has been shown that a nonnegligible amount of the Alfv\'en wave energy driven at the photosphere can reach coronal altitudes \citep{soler2017,soler2019,soler2021}. A thorough examination of the theoretical properties of magnetohydrodynamic waves in the stratified solar atmosphere is provided in \citet{cally2024}.

The theoretical study of Alfv\'en wave reflection in the stratified solar atmosphere dates back to the pioneering works of \citet{ferraro1954} and \citet{ferraro1958}. Since then, an extensive body of literature on Alfv\'en wave reflection has been produced, including  works such as \citet{hollweg1978,leroy1980,Hollweg1981,Leer1982,Zhugzhda1982,hollweg1984,An1989,musielak1992,velli1989,velli1993,campos1998,Cranmer2005,cally2012,soler2017}, among others. A key property of unidirectional Alfv\'en waves in a uniform plasma is the equipartition of kinetic and magnetic energy in the wave perturbations, which allows the energy flux to be expressed as proportional solely to the kinetic energy density. However, in the presence of wave reflection, equipartition of kinetic and magnetic energy no longer holds \citep[see, e.g.,][]{An1989,campos1998,Cranmer2005,Hollweg2007}. Consequently, the classic expression that relates the Alfv\'en wave energy flux to the kinetic energy density is not generally applicable, as it assumes unidirectional propagation and energy equipartition. This limitation has significant implications for accurately estimating the energy content of observed waves, although its importance has largely gone unnoticed by the community.

This paper deals with the role of longitudinal stratification, although transverse nonuniformity can also be important. \citet{goossens2013} demonstrated that local equipartition between kinetic and magnetic energy is not satisfied for  kink waves propagating in a flux tube, though global equipartition is obtained when energies are integrated across the flux tube. They also showed that the radial variation of energy density and energy flux leads to an overestimation of the actual energy content by the classic Alfv\'en wave energy flux expression. Furthermore, \citet{vandoorsselaere2014} expanded on this by considering ensembles of flux tubes and highlighted the importance of the filling factor (the ratio of the flux tube volume to the total volume) in accurately determining wave energy content. However, neither \citet{goossens2013} nor \citet{vandoorsselaere2014} considered longitudinal stratification.

The goal of this paper is to emphasize the importance of reflection, induced by longitudinal stratification, on the energy content and energy flux of Alfv\'en waves propagating in the lower solar atmosphere. To achieve this, we deliberately neglect the role of transverse structuring in this study. Section~\ref{sec:equations} presents the basic equations for addressing the problem of linear Alfv\'en waves propagating in a plasma with field-aligned nonuniformity. A simple analytic model is introduced in Section~\ref{sec:simplemodel} to illustrate the effect of wave reflection on monochromatic Alfv\'en waves in an exponentially stratified plasma. Then, Section~\ref{sec:atmosphere} includes a numerical study of Alfv\'en wave propagation in a model of the lower solar atmosphere and Section~\ref{sec:application} discusses how the proposed expression for the energy flux should be  applied using observations. Finally, Section~\ref{sec:conclusion} provides some concluding remarks.


\section{General expressions} \label{sec:equations}

We consider a plasma  with a uniform magnetic field in the $z$-direction as ${\bf B} = B_0 \hat{e}_z$, with $B_0$ constant. The plasma density varies along the direction of the magnetic field, namely $\rho = \rho(z)$. The linearized ideal MHD equations for the discussion of Alfv\'en waves in this configuration are,
\begin{eqnarray}
    \frac{\partial v_\perp}{\partial t} &=& \frac{B_0}{\mu \rho}  \frac{\partial b_\perp}{\partial z}, \label{eq:main1} \\
     \frac{\partial b_\perp}{\partial t} &=& B_0 \frac{\partial v_\perp}{\partial z}, \label{eq:main2}
\end{eqnarray}
where $v_\perp$ and $b_\perp$ are the perpendicular components of the velocity and magnetic field  perturbations, respectively, and $\mu$ is the magnetic permeability. A conservation equation for the wave energy can be obtained by multiplying Equation~(\ref{eq:main1}) by $v_\perp$ and Equation~(\ref{eq:main2}) by $b_\perp/\mu$ and adding the resulting expressions \citep[see, e.g.,][]{leroy1985,walker2005}. We arrive at,
\begin{equation}
    \frac{\partial E}{\partial t} + \frac{\partial F}{\partial z} = 0, \label{eq:energyconservation}
\end{equation}
where $E = E_k + E_m$ is the total wave energy density, with $E_k$ and $E_m$ the kinetic and magnetic energy densities, respectively, and $F$ is the energy flux. The expressions of these quantities are,
\begin{eqnarray}
    E_k &=& \frac{1}{2}\rho  v_\perp^2, \label{eq:ek} \\
    E_m &=& \frac{1}{2\mu} b_\perp^2, \label{eq:em} \\
    F &=& - \frac{B_0}{\mu} v_\perp b_\perp. \label{eq:flux}
\end{eqnarray}
Although the analysis involves linear waves, it is important to note that the wave energies and the energy flux are quadratic in the perturbations.

Equations~(\ref{eq:main1}) and (\ref{eq:main2}) can also conveniently be combined to eliminate either $v_\perp$ or $b_\perp$ and obtain a wave equation for the remaining perturbation alone. The resulting wave equations are,
\begin{eqnarray}
    \frac{\partial^2 v_\perp}{\partial t^2} - v_{\rm A}^2  \frac{\partial^2 v_\perp}{\partial z^2} &=& 0, \label{eq:wavev} \\
     \frac{\partial^2 b_\perp}{\partial t^2} - v_{\rm A}^2  \frac{\partial^2 b_\perp}{\partial z^2} &=& \frac{\partial v_{\rm A}^2}{\partial z}   \frac{\partial b_\perp}{\partial z} , \label{eq:waveb}
\end{eqnarray}
where $v_{\rm A} = B_0 / \sqrt{\mu\rho}$ is the Alfv\'en speed. We note that  $v_{\rm A} = v_{\rm A}(z)$ owing to the dependence with the density. It is evident that $v_\perp$ and $b_\perp$ satisfy different  equations in a plasma in which $v_{\rm A}$ varies along the direction of the magnetic field. Particularly, the term on the right-hand side of Equation~(\ref{eq:waveb}) acts as a source/sink term for $b_\perp$. This is related to the know result that Alfv\'en waves undergo continuous reflection in the presence of field-aligned nonuniformity \citep[see, e.g.,][]{ferraro1954,hollweg1978,leroy1980,Leer1982}. Let us imagine a wave that  propagates initially towards the positive $z$-direction, i.e., a forward-propagating wave, in a medium where the density decreases with $z$, so that $v_{\rm A}$ increases with $z$. The term on the right-hand side of Equation~(\ref{eq:waveb}) would produce the decrease of the amplitude of $b_\perp$  with $z$.  The amplitude decrease is due to continuous reflection owing to the density gradient because there is no energy dissipation in  ideal MHD. A backward-propagating wave would  be generated by the reflection process, extracting some energy from the forward-propagating wave. If the net energy flux, $F$, is constant, then the amplitude of $v_\perp$ should increase with $z$ according to Equation~(\ref{eq:flux}). The different behavior of $v_\perp$ and $b_\perp$ must have an impact on the  kinetic and magnetic energy content of the wave.

Although one should expect that Alfv\'en wave reflection generally occurs in a plasma with longitudinal stratification, it may be inhibited for certain specific profiles of 
$v_{\rm A}$. As demonstrated by \citet{ruderman2013}, the one-dimensional wave equation can be transformed into the Klein-Gordon equation with constant coefficients when the wave phase speed varies linearly or quadratically with the spatial variable, enabling non-reflective wave propagation. \citet{hansen2012} showed that a one-dimensional wave equation with an exponentially varying phase speed is isomorphic to a two-dimensional wave equation with a constant phase speed \cite[a broader generalization is provided in][]{cally2012}. However, according to these authors, the waves exhibit wakes, which might be interpreted as a form of  reflection.

As an alternative to the concept of continuous reflection, several studies \citep[e.g.,][]{moore1991,musielak1992,Musielak1995} have investigated the existence of cutoff frequencies that depend on the gradient of $v_{\rm A}$, defining specific locations where Alfvén waves are reflected. This approach relies on transforming the wave equation into the Klein-Gordon equation. \citet{perera2015} showed that there is no single, definitive cutoff frequency. Instead, multiple cutoff frequencies can be defined based on the choice of the perturbation used to describe the waves and the method applied to determine the cutoffs. This variability renders the cutoff-frequency interpretation somewhat less compelling than the concept of continuous reflection \citep[see also the comments in][]{velli1993,Tsap2012}.

In the case of a uniform plasma with   $\rho = \rho_0$ and $v_{\rm A} = v_{\rm A,0}$, both constant,  $v_\perp$ and $b_\perp$ satisfy  the same wave equation because the right-hand side term of Equation~(\ref{eq:waveb}) vanishes. There is no reflection when the Alfv\'en speed is uniform and purely unidirectional waves can propagate. In this case, $v_\perp$ and $b_\perp$ are related through the Wal\'en relations \citep{walen1944} as,
\begin{equation}
    v_\perp = \mp \frac{b_{\perp}}{\sqrt{\mu\rho_0}}, \label{eq:walen}
\end{equation}
with the $-$ and $+$ signs corresponding to forward-propagating and backward-propagating  waves, respectively. Substituting Equation~(\ref{eq:walen}) into Equation~(\ref{eq:ek}), we find the well-known result that there is always equipartition of kinetic and magnetic energy, i.e., $E_k = E_m$, so that
\begin{equation}
    E = 2 E_k = 2 E_m. \label{eq:equipartition}
\end{equation}
In addition, from Equation~(\ref{eq:flux}) the wave energy flux adopts the familiar expression,
\begin{equation}
    F = \pm v_{\rm A,0} E, \label{eq:flux2}
\end{equation}
where the $+$ and $-$ signs correspond to forward-propagating and backward-propagating  waves. Equation~(\ref{eq:flux2}) plainly shows that the wave energy, $E$, is simply transported  along the magnetic field at the Alfv\'en speed, $v_{\rm A,0}$, following the direction of  wave propagation. Since there is energy equipartition, Equation~(\ref{eq:flux2}) is frequently presented in an alternative form involving the kinetic energy density, namely
\begin{equation}
    F = \pm v_{\rm A,0} 2E_k = \pm v_{\rm A,0}\rho_0 v_\perp^2, \label{eq:flux3}
\end{equation}
where the velocity perturbation, $v_\perp$, is a quantity that can be observed.

Now, an important warning is needed:  the expressions of the energy flux in Equations~(\ref{eq:flux2}) and (\ref{eq:flux3}) are not general. Equation~(\ref{eq:flux2}) only applies to the  case of unidirectional propagation, while Equation~(\ref{eq:flux3}) further adds the requirement of kinetic and magnetic energy equipartition.  Both conditions can only be realized in a plasma with uniform density. In the presence of longitudinal  stratification, the situation is completely different  because of continuous  reflection.

For our discussion of how the continuous reflection affects the waves energy flux and energy content, it is  useful to define the auxiliary Els\"asser variables \citep{Elsasser1950} as,
\begin{eqnarray}
      Z^{\uparrow} &=& v_{\perp} - \frac{b_{\perp}}{\sqrt{\mu\rho}}, \\
    Z^{\downarrow} &=& v_{\perp} + \frac{b_{\perp}}{\sqrt{\mu\rho}}. 
\end{eqnarray}
Here $Z^\uparrow$ and $Z^\downarrow$ represent forward-propagating and backward-propagating Alfv\'en waves with respect to the direction of the background magnetic field. It is easy to see that the Wal\'en relations (Equation~(\ref{eq:walen})) can be obtained from the Els\"asser variables by considering unidirectional propagation. With the help of the Els\"asser variables,  the  energy flux in Equation~(\ref{eq:flux}) can  be decomposed as 
\begin{equation}
    F = F^\uparrow - F^\downarrow,
\end{equation}
where $F^\uparrow$ and $F^\downarrow$ denote the energy fluxes associated with forward and backward  waves, namely
\begin{eqnarray}
    F^\uparrow &=& \frac{B_0}{4} \sqrt{\frac{\rho}{\mu}}Z^{\uparrow}Z^{\uparrow}, \label{eq:fluxupgen0}\\ 
    F^\downarrow &=& \frac{B_0}{4} \sqrt{\frac{\rho}{\mu}}Z^{\downarrow}Z^{\downarrow}.\label{eq:fluxdowngen0}
\end{eqnarray}
We note that these fluxes are positive by construction. If the process of continuous reflection causes both  forward-propagating and backward-propagating Alfv\'en waves to be simultaneously present, then the net energy flux is the superposition of the two fluxes in opposite directions. Now, we use again the Els\"asser variables in Equations~(\ref{eq:ek}) and (\ref{eq:em}) to check if this simple rule applies for the kinetic and magnetic energies. We obtain the expressions,
\begin{eqnarray}
     E_k &=& \frac{1}{8}\rho \left( Z^{\uparrow}Z^{\uparrow} + Z^{\downarrow}Z^{\downarrow} + 2 Z^{\uparrow}Z^{\downarrow} \right), \label{eq:ekelssaser} \\
    E_m &=& \frac{1}{8}\rho \left( Z^{\uparrow}Z^{\uparrow} + Z^{\downarrow}Z^{\downarrow} - 2 Z^{\uparrow}Z^{\downarrow} \right).\label{eq:emelssaser}
\end{eqnarray}
The total energy, $ E = E_k + E_m$, is,
\begin{equation}
     E = \frac{1}{4}\rho \left( Z^{\uparrow}Z^{\uparrow} + Z^{\downarrow}Z^{\downarrow}\right) = E^{\uparrow} + E^{\downarrow}, \label{eq:totenergy}
\end{equation}
where $E^{\uparrow}$ and $E^{\downarrow}$ denote the total energies associated with the forward and backward waves, respectively. The total energy is indeed the sum of the total energies of the two waves, but the expressions of $E_k$ and $E_m$ contain a cross term proportional to $Z^{\uparrow}Z^{\downarrow}$ that makes it impossible to express them as the sum of the kinetic/magnetic energies of the individual waves \citep[see also][]{Cranmer2005}. Another consequence of the presence of this cross term is that, generally, $E_k \neq E_m$. In other words, there is no equipartition of kinetic and magnetic energies when both forward-propagating and backward-propagating Alfv\'en waves are present due to continuous reflection in a stratified plasma.

Equation~(\ref{eq:totenergy}) can be combined with Equations~(\ref{eq:fluxupgen0}) and (\ref{eq:fluxdowngen0}) to find an expression that relates the  energy flux to the energy density as
\begin{equation}
   F = \mathcal{R}\,  v_{\rm A}\, E, \label{eq:netenergy}
\end{equation}
with $\mathcal{R} \in [-1,1]$ the reflection factor given by,
\begin{equation}
  \mathcal{R} =   \frac{F^\uparrow - F^\downarrow}{F^\uparrow + F^\downarrow}. \label{eq:reffactor}
\end{equation}
The reflection factor is $ \mathcal{R}=1$ for unidirectional forward-propagating waves ($F^\uparrow \neq 0$ and $F^\downarrow = 0$), $ \mathcal{R}=-1$ for unidirectional backward-propagating waves ($F^\uparrow = 0$ and $F^\downarrow \neq 0$), and $ \mathcal{R}=0$ for standing waves ($F^\uparrow = F^\downarrow$). In any other scenario, $\mathcal{R}$ varies between $-1$ and $1$. The ratio $F/E$ informs about the effective velocity of net energy propagation by the Alfv\'en waves. When both forward and backward fluxes are present, Equation~(\ref{eq:netenergy}) evidences that this velocity is smaller than the local Alfv\'en speed because $|\mathcal{R}| < 1$. Therefore, the process of continuous reflection also causes  the velocity at which the Alfv\'en waves transport net energy to slow down and become sub-Alfv\'enic \citep[see also][]{Zhugzhda1982}.

Furthermore, the relation between the kinetic energy and the total energy is,
\begin{equation}
    E_k = \mathcal{N}\, E, \label{eq:energiesrelation}
\end{equation}
with $\mathcal{N} \in [0,1]$ the non-equipartition factor given by,
\begin{equation}
  \mathcal{N} = \frac{1}{2} + {\rm sign}\left(Z^{\uparrow}Z^{\downarrow}\right) \frac{\sqrt{F^\uparrow  F^\downarrow}}{F^\uparrow + F^\downarrow}. \label{eq:equifactor}
\end{equation}
Kinetic and magnetic energy equipartition corresponds to $\mathcal{N} = 1/2$ and can only occur when either $F^\uparrow$ or  $F^\downarrow$ vanishes, i.e., for unidirectional propagation. Equipartition is not possible when both $F^\uparrow$ and  $F^\downarrow$ are present. The relative weights of the two energies depend on the sign of the cross term $Z^{\uparrow}Z^{\downarrow}$, so that $E_k > E_m$ when $Z^{\uparrow}Z^{\downarrow}>0$ and $E_k < E_m$ when $Z^{\uparrow}Z^{\downarrow}<0$. Accordingly, Equation~(\ref{eq:netenergy}) can be recast as,
\begin{equation}
    F = \frac{\mathcal{R}}{\mathcal{N}}\,v_{\rm A}\, E_k = \frac{\mathcal{R}}{2\mathcal{N}}\,v_{\rm A}\, \rho\, v_\perp^2.  \label{eq:netenergy2}
\end{equation}
 It is worth noting that if the entire energy content is composed solely of magnetic energy ($\mathcal{N} \to 0$), the expression in Equation~(\ref{eq:netenergy2}) becomes invalid. Anyway, there would be no observable   velocity perturbation in this particular case.

In the presence of longitudinal stratification, it is misleading to use the expression of the energy flux for unidirectional Alfv\'en waves in a uniform plasma. Based on the above discussion,  Equations~(\ref{eq:flux2}) and (\ref{eq:flux3}) should  be  generalized in a stratified plasma as Equations~(\ref{eq:netenergy}) and (\ref{eq:netenergy2}), respectively, with Equations~(\ref{eq:flux2}) and (\ref{eq:flux3}) being recovered when $\mathcal{R} = \pm 1$ and $\mathcal{N} = 1/2$. The deviation of the correction factors $\mathcal{R}$ and $\mathcal{N}$ from their canonical values for unidirectional Alfv\'en waves depends on the strength of wave reflection. This, in turn, is determined by the specific Alfv\'en speed profile and the frequency or wavelength of the Alfv\'en waves. Generally, for arbitrary profiles, both correction factors would also vary as functions of position.

\section{Analytic model} \label{sec:simplemodel}

Here, a simple model is used  to illustrate the effects of continuous reflection on Alfv\'en waves propagating in a plasma with longitudinal  stratification. For monochromatic Alfv\'en waves with frequency $\omega$, the perturbations can be expressed as proportional to $\exp\left(- i\omega t \right)$. We note that no specific prescription for the temporal dependence was assumed in Section~\ref{sec:equations}. Now, the wave equation for $v_\perp$ (Equation~(\ref{eq:wavev})) becomes,
\begin{equation}
    \frac{\partial^2 v_\perp}{\partial z^2} + \frac{\omega^2}{v_{\rm A}^2(z)} v_\perp = 0, \label{eq:wavev2}
\end{equation}
while the relation between $b_\perp$ and $v_\perp$ is
\begin{equation}
    b_\perp = \frac{iB_0}{\omega} \frac{\partial v_\perp}{\partial z}. \label{eq:relbperp}
\end{equation}
Let us assume that the $z$-direction aligns with the vertical direction and consider a medium in which the Alfv\'en speed increases exponentially with height as,
\begin{equation}
    v_{\rm A}(z) = v_{\rm A,0} \exp\left( z/\Lambda\right), \label{eq:profile}
\end{equation}
where $v_{\rm A,0}$ is the Alfv\'en speed at the reference height $z=0$ and $\Lambda$ is an arbitrary scale height. The choice of an exponential profile allows for analytic solutions  \citep[see, e.g.,][]{ferraro1958,hollweg1978}. In a stratified plasma, the quantity 
\begin{equation}
    \lambda(z) = \frac{v_{\rm A}(z)}{\omega},
\end{equation}
plays the role of a local wavelength.  We perform the change of variable,
\begin{equation}
    \xi(z) = \frac{\Lambda}{\lambda(z)} = \frac{\Lambda}{\lambda_0}\exp\left(- z/\Lambda\right) = \xi_0 \exp\left(- z/\Lambda\right),
\end{equation}
where $\lambda_0 =  v_{\rm A,0}/\omega$ and $\xi_0 = \Lambda/\lambda_0$. The solution to  Equation~(\ref{eq:wavev2}) in the new variable is,
\begin{equation}
    v_\perp(\xi) = C_1 H_0^{(1)}\left( \xi \right)  +C_2 H_0^{(2)}\left( \xi\right), \label{eq:solwave}
\end{equation}
where $H_0^{(1)}$ and $H_0^{(2)}$ are the Hankel functions of the first and second kinds of order 0, respectively. In turn, the expression for $b_\perp$ using Equation~(\ref{eq:relbperp}) is
\begin{equation}
    b_\perp (\xi) =  \frac{i B_0}{\Lambda\omega} \xi \left( C_1 H_1^{(1)}\left( \xi \right)  +C_2 H_1^{(2)}\left( \xi\right)\right), \label{eq:solwave2}
\end{equation}
where now the Hankel functions are of order 1.  Equations~(\ref{eq:solwave}) and (\ref{eq:solwave2}) depend on the complex constants $C_1$ and $C_2$. In practical scenarios, these constants would be determined by the mechanism that drives the waves and the imposed boundary conditions. However, in the following analysis, we deliberately avoid specifying any particular conditions for these constants to ensure the discussion remains as general as possible. This approach allows the effects discussed below to be attributed solely to the influence of stratification.

Let us compute the wave energy flux using the expressions of $v_\perp$ and $b_\perp$. Due to the harmonic time dependence, Equation~(\ref{eq:flux}) must be averaged over one complete wave cycle to retain the net contribution  \citep[see][]{walker2005}, namely
\begin{equation}
   \left< F \right> = - \frac{B_0}{2\mu} {\rm Re}\left[ v_\perp b_\perp^* \right], \label{eq:fluxav} 
\end{equation}
where the asterisk indicates the complex conjugate and we used the notation $\left< \cdots \right>$ to denote the temporal average. After some algebra, the following expression for the energy flux can be obtained \citep[see, e.g.,][]{hollweg1984,cally2024},
\begin{equation}
     \left< F \right> = \frac{B^2_0}{\mu\pi\Lambda\omega} \left(  \left|C_2 \right|^2 - \left|C_1 \right|^2\right),\label{eq:fluxav2} 
\end{equation}
which is independent of height in this model without dissipation. Equation~(\ref{eq:fluxav2}) has a positive contribution proportional to $\left|C_2 \right|^2$ and a negative contribution proportional to $\left|C_1 \right|^2$. This led some authors to associate the  function $H^{(2)}$ to upward-propagating waves and the function $H^{(1)}$ to downward-propagating waves, with the constants $C_1$ and $C_2$ playing the role of the waves amplitudes. Although conceptually useful, this direct association is incomplete. To show that, we compute the Els\"asser variables, namely
\begin{eqnarray}
    Z^\uparrow &=& C_1 \left( H_0^{(1)}\left( \xi \right) - i H_1^{(1)}\left( \xi \right) \right) \nonumber \\ && + C_2 \left( H_0^{(2)}\left( \xi \right) - i H_1^{(2)}\left( \xi \right) \right), \\
    Z^\downarrow &=& C_1 \left( H_0^{(1)}\left( \xi \right) + i H_1^{(1)}\left( \xi \right) \right) \nonumber \\ && + C_2 \left( H_0^{(2)}\left( \xi \right) + i H_1^{(2)}\left( \xi \right) \right). 
\end{eqnarray}
Clearly, both Els\"asser variables contain terms involving $C_1$ and $C_2$.  It is impossible for one Els\"asser variable to vanish for all $\xi$ while the other remains nonzero \citep[see also][]{velli1989}. This implies that no combination of the constants $C_1$ and $C_2$ allows for unidirectional propagation, even if one of them is set to zero. So, how can the energy flux take the simple form given by Equation~(\ref{eq:fluxav2})? To address this, we compute the averaged upward and downward fluxes using the Els\"asser variables, namely
\begin{eqnarray}
    \left< F^\uparrow \right> &=& \frac{B_0}{8} \sqrt{\frac{\rho}{\mu}}{\rm Re}\left[Z^{\uparrow}Z^{\uparrow*}\right],\\ 
     \left< F^\downarrow \right> &=& \frac{B_0}{8} \sqrt{\frac{\rho}{\mu}}{\rm Re}\left[Z^{\downarrow}Z^{\downarrow*}\right].
\end{eqnarray}
 We obtain,
\begin{eqnarray}
    \left< F^\uparrow \right> &=& \frac{B^2_0}{2\mu\pi\Lambda\omega} \left(   \left|C_2 \right|^2 - \left|C_1 \right|^2 \right. \nonumber \\
    && + \left. \left(  \left|C_1 \right|^2 + \left|C_2 \right|^2\right)  \phi(\xi)\right), \label{eq:fluxupgen}\\ 
     \left< F^\downarrow \right> &=& \frac{B^2_0}{2\mu\pi\Lambda\omega} \left(   \left|C_1 \right|^2 - \left|C_2 \right|^2 \right. \nonumber \\
    && + \left. \left(  \left|C_1 \right|^2 + \left|C_2 \right|^2\right)  \phi(\xi) \right), \label{eq:fluxdowngen}
\end{eqnarray}
where $\phi(\xi) = \phi_0(\xi) +\phi_1(\xi)$ with
\begin{equation}
    \phi_n(\xi) = G_n(\xi) + \frac{{\rm Re}\left[C_1 C_2^*\right]L_n(\xi)-{\rm Im}\left[C_1 C_2^*\right]M_n(\xi)}{ \left|C_1 \right|^2 + \left|C_2 \right|^2}, 
\end{equation}
 where
\begin{eqnarray}
    G_n(\xi) &=& \frac{\pi\xi}{4} \left( J_n(\xi)^2 + Y_n(\xi)^2 \right),\\
    L_n(\xi) &=&  \frac{\pi\xi}{2} \left( J_n(\xi)^2  - Y_n(\xi)^2 \right),\\
    M_n(\xi) &=&  \pi\xi  J_n(\xi)Y_n(\xi).
\end{eqnarray}
In these expressions, $J_n$ and $Y_n$ are the Bessel functions of the first and second kinds of order $n$, respectively, with $n=$~0 and 1. Both upward and downward fluxes contain terms with $C_1$ and $C_2$. For example, if $ C_2 = 0 $, $ \left< F^\uparrow \right> $ remains nonzero due to  the terms with $ \left|C_1 \right|^2 $, and similarly, $ \left< F^\downarrow \right> $ remains nonzero if $ C_1 = 0 $. In addition, the two fluxes are not constant on position but have an oscillatory component due to the function $\phi(\xi)$.  The net flux of Equation~(\ref{eq:fluxav2}) is simply recovered by subtracting the  upward and downward fluxes as $\left< F \right> = \left< F^\uparrow \right> - \left< F^\downarrow \right>$. Then, as similarly discussed by \citet{velli1989}, the upward and downward fluxes interact coherently, their oscillatory parts cancel out, and the net flux becomes  constant.

It is instructive the consider the limit that the scale height is much larger than the local wavelength, i.e., we take $\xi \gg 1$. In other words, we assume that the Alfv\'en speed gradient is sufficiently gentle. Asymptotic expansion of the Hankel functions for large arguments \citep[see expressions 10.2.5 and 10.2.6 in][]{NIST:DLMF} provides   approximate expressions for $v_\perp$ and $b_\perp$, namely
\begin{eqnarray}
     v_\perp(\xi) &\approx &\frac{1}{\sqrt{\xi}} \left( \tilde{C}_1 \exp\left( i \xi\right) + \tilde{C}_2 \exp\left( -i \xi\right)\right), \\
     b_\perp(\xi) &\approx & \frac{B_0}{\Lambda\omega}\sqrt{\xi} \left( \tilde{C}_1 \exp\left( i \xi\right) - \tilde{C}_2 \exp\left( -i\xi\right)\right),
\end{eqnarray}
where $\tilde{C}_1 = \sqrt{2/\pi}e^{-i\pi/4}C_1$ and $\tilde{C}_2 = \sqrt{2/\pi}e^{i\pi/4}C_2$. The Els\"asser variables become,
\begin{eqnarray}
    Z^\uparrow &\approx & 2C_2 \sqrt{\frac{2}{\pi\xi}} \exp\left( -i(\xi-\pi/4)\right), \\
    Z^\downarrow &\approx & 2C_1 \sqrt{\frac{2}{\pi\xi}} \exp\left( i(\xi-\pi/4)\right). 
\end{eqnarray}
Now, unidirectional propagation is possible when either $C_1$ or $C_2$ vanishes.  However, this  is an artifact of the asymptotic approximation and does not happen outside this limit. Concerning the energy fluxes,  we find that in this approximation \citep[see expressions 10.7.8 in][]{NIST:DLMF},
\begin{equation}
    \phi_n(\xi) \approx \frac{1}{2} + (-1)^n \frac{{\rm Re}\left[C_1 C_2^*\right]\sin(2\xi)+{\rm Im}\left[C_1 C_2^*\right]\cos(2\xi)}{ \left|C_1 \right|^2 + \left|C_2 \right|^2}, \label{eq:phiapp}
\end{equation}
so that $\phi(\xi) = \phi_0(\xi) +\phi_1(\xi)\approx  1$ and the upward and downward fluxes simply become,
\begin{eqnarray}
    \left< F^\uparrow \right> & \approx & \frac{B^2_0}{\mu\pi\Lambda\omega}\left|C_2 \right|^2 ,\\ 
     \left< F^\downarrow \right> & \approx & \frac{B^2_0}{\mu\pi\Lambda\omega} \left|C_1 \right|^2.
\end{eqnarray}
Both fluxes are constant and, when subtracted, provide exactly the same net flux of Equation~(\ref{eq:fluxav2}) obtained in the general case.

Therefore, the association of $H^{(2)}$ with upward-propagating waves and $H^{(1)}$ with downward-propagating waves can only be done in the asymptotic limit of large arguments (gentle Alfv\'en speed gradient).  It would be more accurate to state that $H^{(2)}$ is predominantly upward-propagating, while $H^{(1)}$ is predominantly downward-propagating. In general,  it is not  possible to explicitly disentangle the two directions of propagation from the  complete mathematical solution  \citep[see also][]{Zhugzhda1982}.  The implication of this result is that the perturbations $v_\perp$ and $b_\perp$ are necessarily a superposition of upward-propagating  and downward-propagating disturbances, regardless of the choice of the constants $C_1$ and $C_2$.

Next, we  study how the kinetic and magnetic   energies  vary with position. Equations~(\ref{eq:ek})--(\ref{eq:em})  averaged over one complete wave cycle yield,
\begin{eqnarray}
   \left< E_k \right> &=& \frac{1}{4}\rho v_\perp v_\perp^*, \label{eq:ekav} \\
   \left< E_m \right> &=& \frac{1}{4\mu}  b_\perp b_\perp^*. \label{eq:emav}
\end{eqnarray}
 Using the previously obtained solutions, we arrive at
\begin{eqnarray}
   \left< E_k \right> &=& \frac{B^2_0}{\mu\pi\Lambda^2\omega^2}  \left(   \left|C_1 \right|^2 + \left|C_2 \right|^2 \right) \phi_0(\xi)\, \xi, \label{eq:ekav2} \\
   \left< E_m \right> &=& \frac{B^2_0}{\mu\pi\Lambda^2\omega^2}  \left(   \left|C_1 \right|^2 + \left|C_2 \right|^2 \right) \phi_1(\xi)\, \xi. \label{eq:emav2} 
\end{eqnarray}
Equations~(\ref{eq:ekav2}) and (\ref{eq:emav2}) evidence that energy equipartition is not satisfied globally and that the kinetic and magnetic energy contents of the perturbations are oscillatory due the functions $\phi_0(\xi)$ and $\phi_1(\xi)$. Depending on position, it is possible that one type of energy dominates the  energy content of the wave perturbations. However, energy equipartition may be possible at specific positions where $\phi_0(\xi)$ and $\phi_1(\xi)$ are equal. In addition, both energies are proportional to $\xi$, which means that they decrease exponentially with height in the original variable, $z$. The averaged total energy, $\left< E \right> =  \left< E_k \right> + \left< E_m \right>$, is
\begin{equation}
    \left< E \right> =  \frac{B^2_0}{\mu\pi\Lambda^2\omega^2}  \left(   \left|C_1 \right|^2 + \left|C_2 \right|^2 \right) \phi(\xi)\, \xi,\label{eq:totalenergy}
\end{equation}
which remains oscillatory due to $\phi(\xi)$ and exponentially decreasing with height due to the $\xi$ factor.

To explicitly show how the kinetic and magnetic energy content of the waves depend on height, we revert to the original variable, $z$, and resort to the approximate  forms of $\phi_0(\xi)$ and $\phi_1(\xi)$ in the asymptotic limit of large arguments (Equation~(\ref{eq:phiapp})), which yield simpler expressions that provide clearer insights. Then, Equation~(\ref{eq:totalenergy}) for the total energy  becomes,
\begin{equation}
  \left< E \right> \approx   \left< E \right>_0 \exp(-z/\Lambda), \label{eq:totalenergy2}
\end{equation}
with $\left< E \right>_0$ the total energy at $z=0$ given by,
\begin{equation}
    \left< E \right>_0  = \frac{B^2_0}{\mu\pi\Lambda\omega v_{\rm A,0}}  \left(   \left|C_1 \right|^2 + \left|C_2 \right|^2 \right).
\end{equation}
 The total energy simply decreases exponentially with $z$  because no oscillatory component remains in the asymptotic limit. Since there is no dissipation in this model, the spatial decrease of the total energy is purely caused by  the stratification. In this scenario, the ratios of the kinetic and magnetic energies to the total energy yield,
    \begin{eqnarray}
\frac{\left< E_k \right>}{ \left< E \right>} &\approx &  \frac{1}{2} + \frac{ C_1 C_2}{C_1^2 + C_2^2}\sin\left(\frac{2\Lambda\omega}{v_{\rm A,0}}  \exp(-z/\Lambda)\right),\label{eq:ekav3} \\
\frac{\left< E_m \right>}{ \left< E \right>} & \approx &  \frac{1}{2} - \frac{ C_1 C_2}{C_1^2 + C_2^2}\sin\left(\frac{2\Lambda\omega}{v_{\rm A,0}}  \exp(-z/\Lambda)\right), \label{eq:emav3} 
\end{eqnarray}
where we assumed, for simplicity, that both $C_1$ and $C_2$ are real. Another choice would change the discussion only slightly, as explained later. By comparing these expressions with the general expressions in terms of the Els\"asser variables (Equations~(\ref{eq:ekelssaser}) and (\ref{eq:emelssaser})), it is easy to identify the cross term that results from the superposition of the oppositely propagating waves. This term causes the non-equipartition of the energies and the oscillatory behavior with height. The cross term drops when the total energy is computed by adding the kinetic and magnetic energies. At sufficiently long distances from the reference height, $z=0$, the argument of the $\sin$ function  becomes small and Equations~(\ref{eq:ekav3}) and (\ref{eq:emav3}) can further be approximated by
    \begin{eqnarray}
\frac{\left< E_k \right>}{ \left< E \right>} &\approx &  \frac{1}{2} + \frac{ C_1 C_2}{C_1^2 + C_2^2}\frac{2\Lambda\omega}{v_{\rm A,0}}  \exp(-z/\Lambda),\label{eq:ekav4} \\
\frac{\left< E_m \right>}{ \left< E \right>} & \approx &  \frac{1}{2} - \frac{ C_1 C_2}{C_1^2 + C_2^2}\frac{2\Lambda\omega}{v_{\rm A,0}}  \exp(-z/\Lambda). \label{eq:emav4} 
\end{eqnarray}
If $C_1 C_2 > 0$, these equations predict, first, a predominance of kinetic energy over magnetic energy at large distances and, later, the eventual recovery of energy equipartition when the exponential factor extinguishes for $z/\Lambda \to\infty$. In contrast, if $C_1 C_2 < 0$, it is the magnetic energy, rather than the kinetic energy, that dominates at large distances before equipartition is restored. The differing behavior of kinetic and magnetic energies with height was also addressed by \citet{campos1998}, who reported a predominance of magnetic energy at sufficiently high altitudes. This result is attributable  to a specific choice of the constants $C_1$ and $C_2$.  In the more general case where both constants are complex, the dependence of the energies on height is more complicated even in the asymptotic approximation. However,  it is generally found that the energies are  oscillatory and that the eventual predominance of one form of energy  is possible.


Finally, we present the specific forms of the reflection factor, $ \mathcal{R} $, and the non-equipartition factor, $ \mathcal{N} $, as introduced in Section~\ref{sec:equations}, within the context of this model. The reflection factor, as defined in Equation~(\ref{eq:reffactor}), can be computed with the help of the averaged upward and downward fluxes of Equations~(\ref{eq:fluxupgen}) and (\ref{eq:fluxdowngen}). Alternatively, it can also be computed by relating the net flux (Equation~(\ref{eq:fluxav2})) with the total energy density (Equation~(\ref{eq:totalenergy})). The result is
\begin{equation}
    \mathcal{R} = \frac{\left|C_2 \right|^2 - \left|C_1 \right|^2}{\left|C_1 \right|^2 + \left|C_2 \right|^2}\frac{1}{\phi(\xi)}. \label{eq:rexp}
\end{equation}
Once more, we observe that even when either $C_1$ or $C_2$ vanishes,  the reflection factor satisfies $|\mathcal{R}| < 1$  due to the function $\phi(\xi)$, which  also causes $\mathcal{R}$ to be oscillatory. Non-reflective propagation is only possible in the asymptotic limit where $\phi(\xi)\approx 1$. Since the effective velocity of energy propagation is $\mathcal{R}\,v_{\rm A}$, we conclude that net energy is transported by the waves at a slower velocity  than  the local Alfv\'en speed  \citep[see also][]{Zhugzhda1982}. In turn, the non-equipartition factor can easily be obtained by relating the total and kinetic energies  as  in Equation~(\ref{eq:energiesrelation}), but using Equations~(\ref{eq:ekav2}) and (\ref{eq:totalenergy}). We find,
\begin{equation}
    \mathcal{N} = \frac{\phi_0(\xi)}{\phi(\xi)},
\end{equation}
which anticipates a highly oscillatory nature of this parameter. To obtain  a more explicit form, we resort  to the asymptotic limit and get,
\begin{equation}
    \mathcal{N} \approx \frac{1}{2} + \frac{C_1 C_2}{C_1^2 + C_2^2}\sin\left(\frac{2\Lambda\omega}{v_{\rm A,0}}  \exp(-z/\Lambda)\right), \label{eq:nfactorapp}
\end{equation}
where we again took $C_1$ and $C_2$ as real, for simplicity. The dependence of the non-equipartition factor with height agrees with the above discussion about the oscillatory behavior of the energies. In fact, Equation~(\ref{eq:nfactorapp}) is the same as Equation~(\ref{eq:ekav3}).

Some important results obtained in the presence of longitudinal stratification are generally applicable beyond the limitations of this simple model. There is necessarily a superposition of waves propagating in opposite directions due to continuous reflection.  There is no equipartition of kinetic and magnetic energies in the Alfv\'en wave perturbations. The relative weights of the two energies depend on position. It may even happen that there is a local predominance of one type of energy  in the total energy content and barely no amount of the other type of energy.  The  velocity of net energy propagation in a stratified plasma is generally smaller than the local Alfv\'en speed. The stronger the reflection, the smaller the velocity at which  net energy propagates.


\section{Alfv\'en wave propagation in the lower solar atmosphere} \label{sec:atmosphere}

We study with a numerical model the propagation of Alfv\'en waves in the vertically stratified  lower solar atmosphere. We compare the numerical results with the predictions of the analytic model of Section~\ref{sec:simplemodel}.

\begin{figure}[ht!]
\centering
\plotone{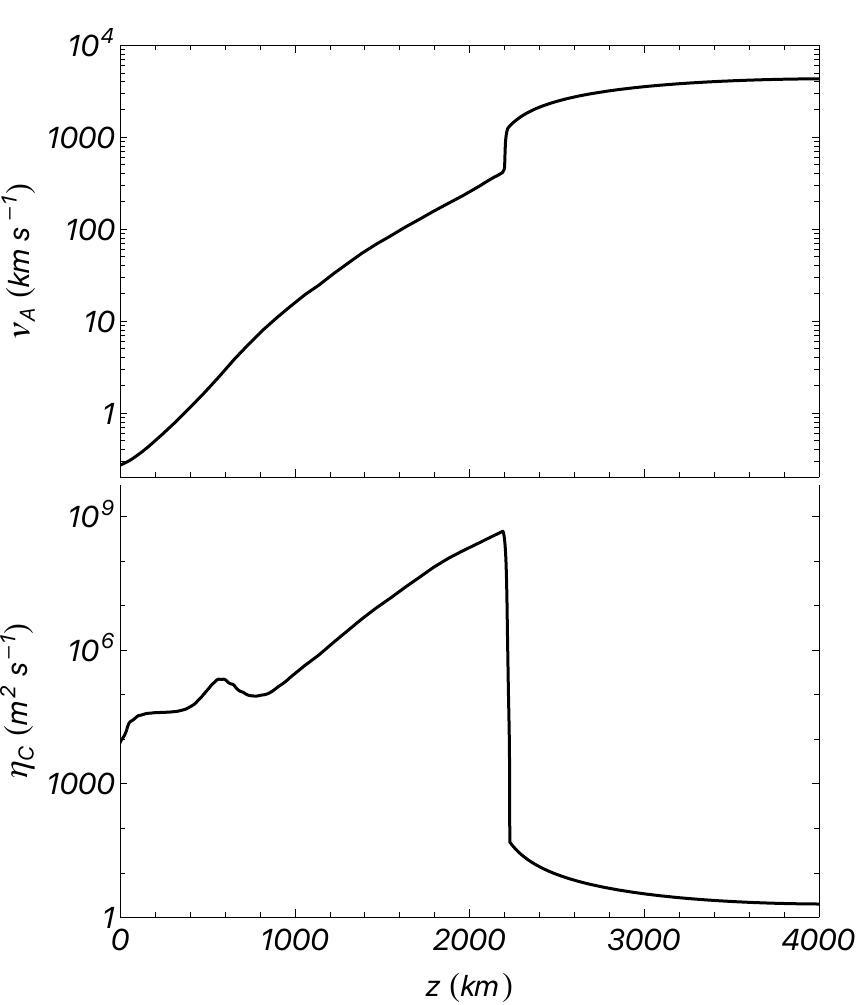}
\caption{Alfv\'en speed  (top) and Cowling's diffusion coefficient (bottom) as functions of height in the model for the lower solar atmosphere. 
\label{fig:model}}
\end{figure}

The  atmospheric model is the same as that   used in \citet{soler2017,soler2019}. It is a static, plane-parallel, and gravitationally stratified model derived from the semi-empirical quiet-Sun model C by \citet{Fontenla1993}. The model spans from the top of the photosphere ($ z = 0$), through the chromosphere and the transition region (located at approximately $ z \approx$ 2,200~km), and extends into the lower corona at $ z =$~4,000~km. A uniform, vertical magnetic field with $B_0 = 50$~G is imposed. The effects of field-line expansion are neglected. In this context, $B_0$ represents the average  magnetic field strength. The role of magnetic expansion on the propagation of Alfv\'en waves was explored in, e.g., \citet{Similon1992,soler2017}. Figure~\ref{fig:model} (top) displays the Alfv\'en speed profile in this model.

\begin{figure*}[ht!]
\centering
\plottwo{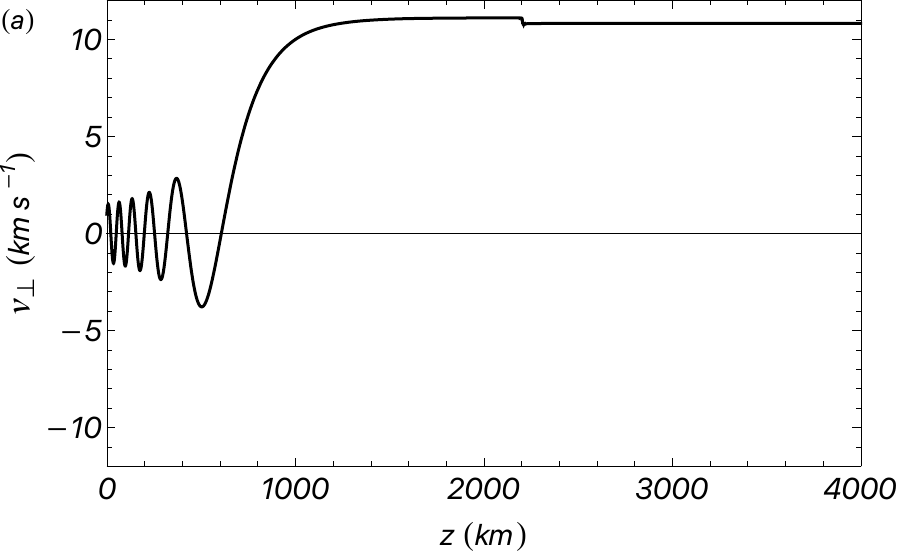}{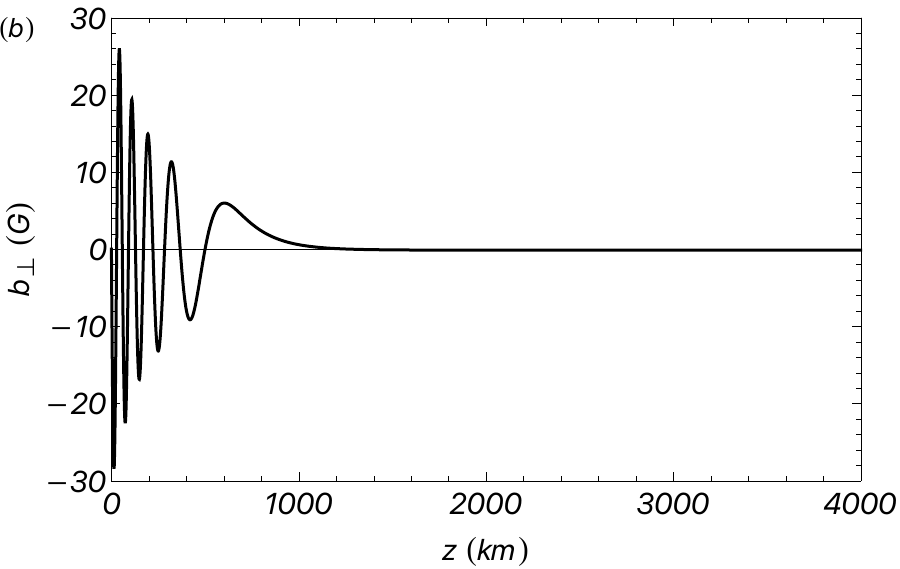}
\plottwo{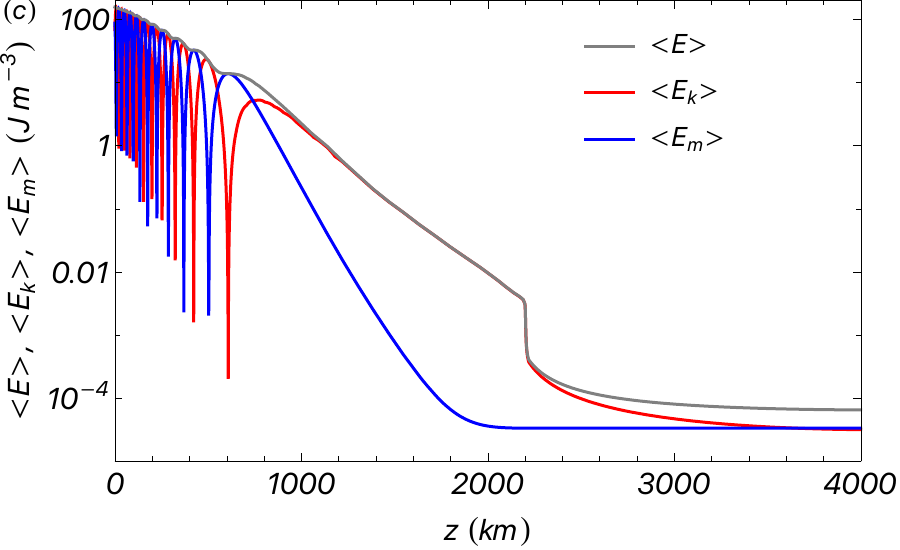}{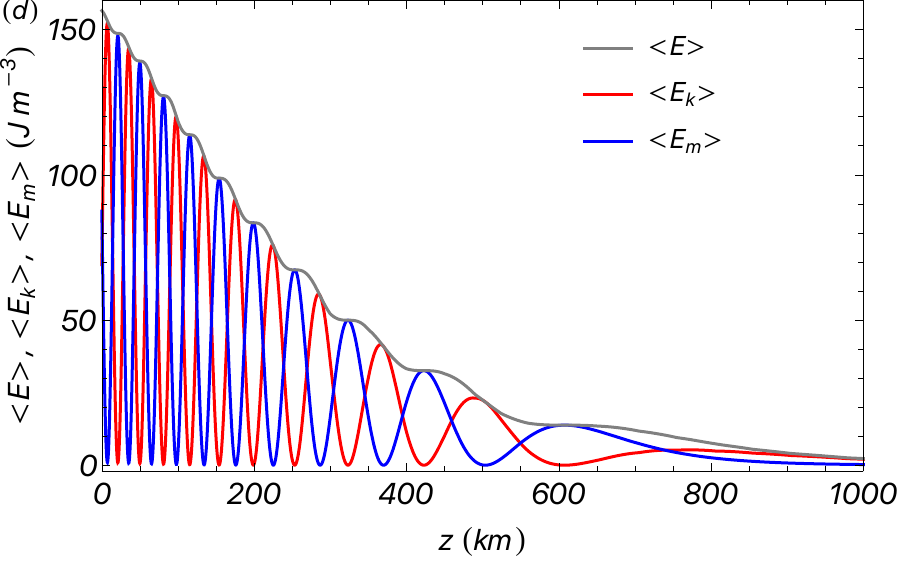}
\plottwo{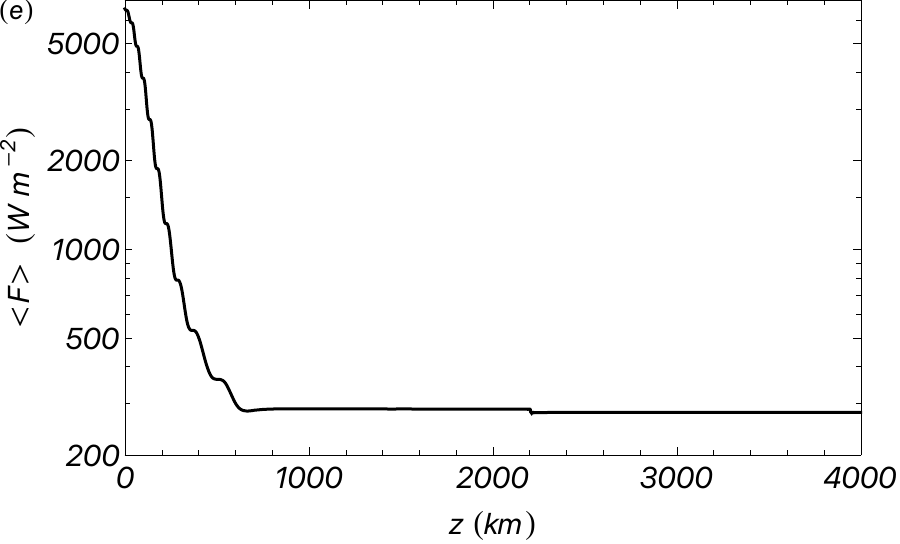}{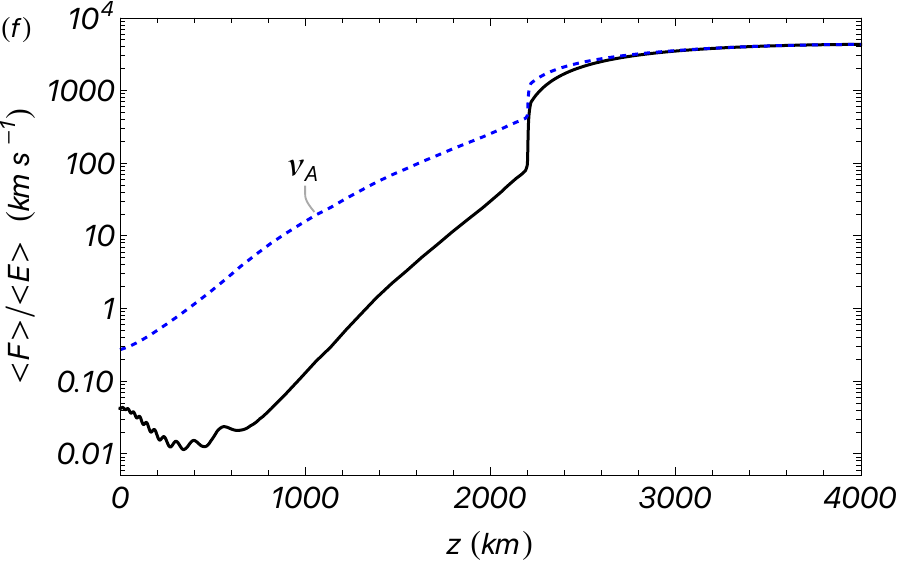}
\caption{Evolution of a 5~mHz Alfv\'en wave as a function of height in the lower solar atmosphere. (a) Velocity perturbation. (b) Magnetic field perturbation. (c) Total,  kinetic, and magnetic energy densities. (d) Same as panel (c), but with a focus on the lower chromosphere at heights below 1,000 km. (e) Net energy flux. (f) Velocity of net energy propagation. The Alfv\'en speed profile has been overplotted in panel (f) for comparison purposes.  Note that the vertical axis in  panels (c), (e), and (f)  is in log-scale.
\label{fig:energy5mHz}}
\end{figure*}

We consider monochromatic Alfv\'en waves that are launched from the photosphere and propagate towards the corona.  Equations~(\ref{eq:main1}) and (\ref{eq:main2}) are numerically integrated in $z$ following the same method as described in \citet{soler2017,soler2019}. The amplitude of the velocity perturbations at the photosphere is set to 1~km~s$^{-1}$, in accordance with the observations of photospheric horizontal motions of \citet{chitta2012}. The upper boundary is perfectly transparent to the incoming waves from below. Partial ionization of the chromospheric plasma causes significant wave dissipation \citep[see][]{soler2024}. To account for this effect, the term,
\begin{equation}
    \frac{\partial}{\partial z} \left(\eta_{\rm C} \frac{\partial b_\perp}{\partial z}  \right),
\end{equation}
 is added to the right-hand side of Equation~(\ref{eq:main2}), with $\eta_{\rm C}$ the coefficient of Cowling's diffusion. Figure~\ref{fig:model} (bottom) displays the variation of $\eta_{\rm C}$ with height in the model. Details on the computation of $\eta_{\rm C}$ can be found in, e.g.,  \citet{soler2015} and are not given here for the sake of simplicity.

Figure~\ref{fig:energy5mHz}(a) and (b) show the velocity and magnetic field perturbations, respectively, as functions of height for an Alfv\'en wave with a frequency of 5~mHz. As expected, the amplitude of $b_\perp$ decreases with height, while that of $v_\perp$ increases. In addition, due to the rapid increase in the Alfv\'en speed, the wavelength increases significantly with height, causing the $ v_\perp $ profile to lose its oscillatory appearance and appear essentially flat at upper altitudes. Wave reflection is significant and causes accumulation of transverse magnetic flux at lower heights, which substantially raises the amplitude of $b_\perp$. This might introduce nonlinear effects on the wave evolution, which are not included in these computations. 

Figure~\ref{fig:energy5mHz}(c) displays the kinetic, magnetic, and total energy densities as functions of height. The total energy decreases rapidly with height. In the lower chromosphere, at heights below approximately 1,000~km, the energy content exhibits significant spatial variability, with alternating dominance between kinetic and magnetic energy. This is more clearly observed in Figure~\ref{fig:energy5mHz}(d), which provides a close-up view of the lower heights.  Conversely, higher up in the middle and upper chromosphere, kinetic energy strongly predominates. This behavior is caused by the extinction of the magnetic field perturbation with height (see Figure~\ref{fig:energy5mHz}(b)).  Above the transition region, the Alfv\'en speed profile becomes smooth, reducing the significance of continuous wave reflection. As a result, the total energy density remains nearly constant, and kinetic and magnetic energies approach near equipartition when the lower corona is reached at $z=$~4,000~km. These results align well with the predictions of the analytic model.

\begin{figure*}[ht!]
\centering
\plotone{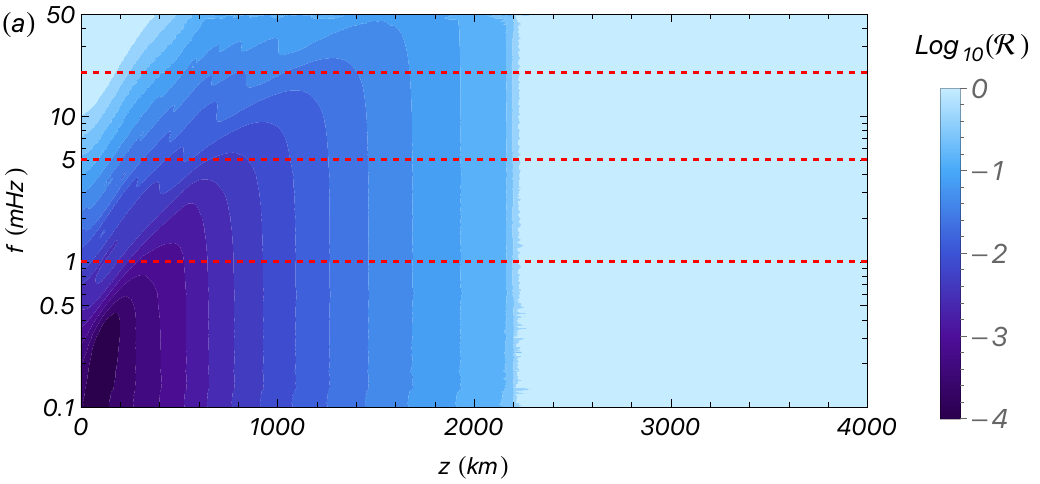}
\plotone{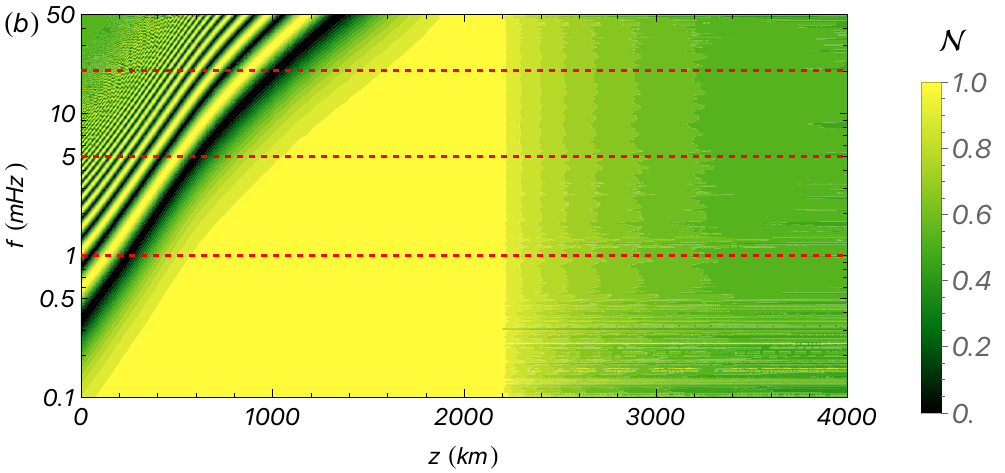}
\caption{Contour plots of the (a) reflection and (b) non-equipartition correction factors to the expression of the energy flux (Equation~(\ref{eq:netenergy2})) as functions of the frequency and height in the lower solar atmosphere. Note that $\log_{10}(\mathcal{R})$ is plotted in panel (a) and that the frequency axis is in log-scale in both panels. The horizontal red dashed lines indicate the results for the three particular frequencies plotted in Figure~\ref{fig:factors}.
\label{fig:factorscontour}}
\end{figure*}

Figure~\ref{fig:energy5mHz}(e) shows the net energy flux, which decreases with height in the lower chromosphere because of Cowling's diffusion and becomes roughly constant at upper heights. The overall decreasing trend of the energy flux in the lower chromosphere is accompanied by small-amplitude oscillations, which result from the interaction between oppositely propagating waves. The  energy flux is always positive, indicating a net upward energy transport. We repeated these computations without including dissipation (results not shown here). The energy density in the ideal case is indistinguishable from the dissipative case. However,  the ideal energy flux remains positive and constant across all heights, as expected. On the other hand, Figure~\ref{fig:energy5mHz}(f) displays the velocity of net energy propagation computed as the ratio of the energy flux to the total energy density.  This velocity is sub-Alfv\'enic in the chromosphere. Above the transition region, it closely approaches the local Alfv\'en speed due to the diminished significance of  reflection. Again, all these findings were anticipated by the analytic model.

\begin{figure}[ht!]
\centering
\plotone{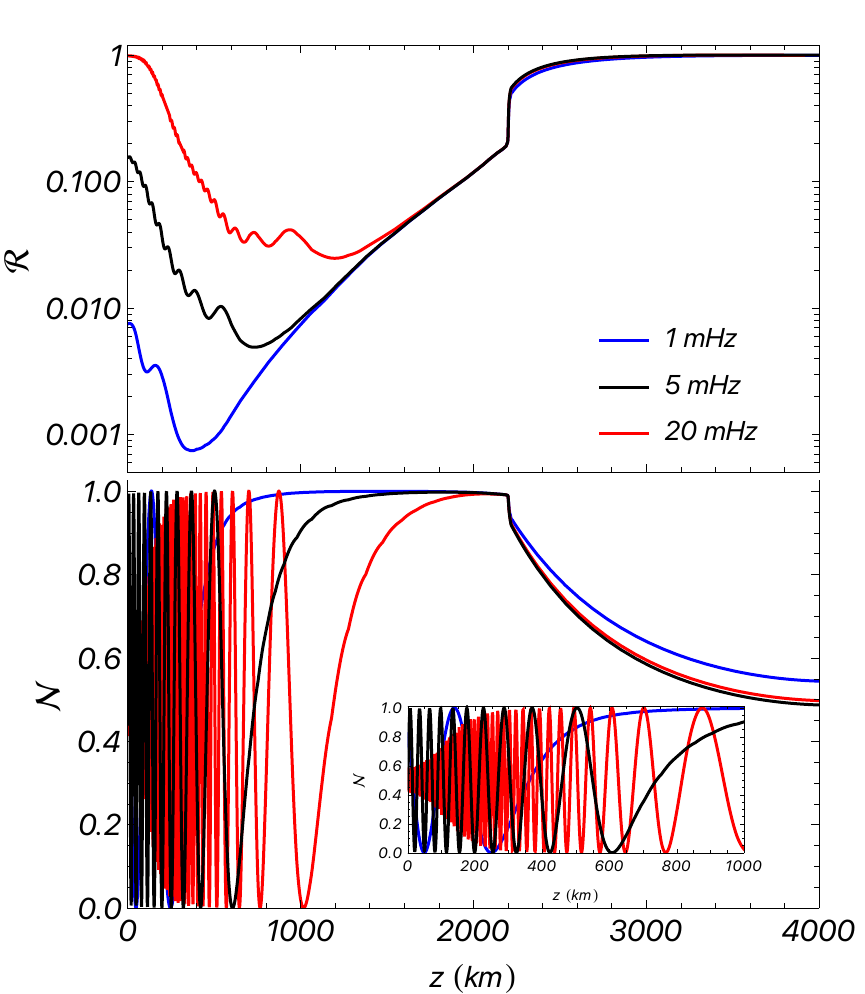}
\caption{Correction factors for reflection (top) and energy non-equipartition (bottom) as functions of height in the lower solar atmosphere for Alfv\'en waves with frequencies of 1~mHz, 5~mHz, and 20~mHz. The inset in the bottom panel shows a detailed view of the non-equipartition factor for heights below 1,000~km. Note that the vertical axis in the top panel is in log-scale.
\label{fig:factors}}
\end{figure}

Further numerical computations have been carried out for waves with frequencies ranging from 0.1~mHz to 50~mHz. These results are used to calculate the reflection, $\mathcal{R}$, and non-equipartition, $\mathcal{N}$, correction factors to the energy flux expression (Equation~(\ref{eq:netenergy2})). Figure~\ref{fig:factorscontour} displays contour plots of these factors in the considered frequency range and as functions of height over the photosphere. Below the transition region, both factors significantly deviate from their canonical values for unidirectional Alfv\'en waves. In the lower chromosphere $\mathcal{R}\ll 1$, specially for low frequencies for which reflection is very relevant. The importance of  reflection diminishes as the frequency increases, although it remains significant in the whole frequency range. It is important to note that $\mathcal{R}$ remains positive at all heights and for all considered frequencies, which indicates that these chromospheric waves produce an net upward energy transport. 

The non-equipartition factor exhibits a highly oscillatory behavior in the lower chromosphere, where the energy content alternates between being predominantly kinetic and predominantly magnetic over very short distances. This oscillatory pattern persists up to a certain height, beyond which the energy content becomes largely kinetic, leading to $\mathcal{N} \approx 1$. However, the extent of the oscillatory region of $\mathcal{N}$ increases with frequency, encompassing nearly the entire chromosphere at high frequencies. In contrast, wave reflection becomes negligible above the transition region. At those heights, both correction factors approach their canonical values, $\mathcal{R} \approx 1$ and $\mathcal{N} \approx 1/2$. To observe in more detail the dependence  of $\mathcal{R}$ and  $\mathcal{N}$ on height, Figure~\ref{fig:factors} displays both factors for Alfv\'en waves with frequencies of 1~mHz, 5~mHz, and 20~mHz. The small values of $\mathcal{R}$ in the entire chromosphere and the oscillatory nature of $\mathcal{N}$ at low altitudes are evident.

Our findings on the non-equipartition of kinetic and magnetic energy in solar atmospheric Alfv\'en waves can be compared to those of \citet{Cranmer2005}, despite differences in the background models, with \citet{Cranmer2005} extending to greater heights. Figure 6 of \citet{Cranmer2005}  displays the ratio of kinetic and magnetic energies as a function of height and wave period. The region where  kinetic and magnetic energies are oscillatory is clearly visible in the lower left part of Figure 6 of \citet{Cranmer2005}, which roughly corresponds to the upper left part of Figure~\ref{fig:factorscontour}(b).  The results are less comparable at larger altitudes, although \citet{Cranmer2005} also obtain the subsequent predominance of kinetic energy and the eventual restoration of energy equipartition for wave periods matching the frequencies examined here.

\section{Application to observations}
\label{sec:application}

Now, we address the applicability of Equation~(\ref{eq:netenergy2}) for the determination of the energy flux of chromospheric waves using observations. For the expression to be used, it is essential to estimate the values of $\mathcal{R}$ and $\mathcal{N}$. The following discussion focuses on explaining how these parameters can be estimated.

The determination of both parameters might be possible if the observed velocity amplitude, $v_\perp$, can somehow be decomposed into the velocity amplitudes of the superimposed upward, $v_\perp^\uparrow$, and downward, $v_\perp^\downarrow$, perturbations. The procedure to perform such a separation  is not discussed here. If the separation is possible, then  $\mathcal{R}$ might be calculated from its definition in Equation~(\ref{eq:reffactor}) as,
\begin{equation}
    \mathcal{R} =  \frac{F^\uparrow - F^\downarrow}{F^\uparrow + F^\downarrow}= \frac{v_\perp^{\uparrow \, 2} - v_\perp^{\downarrow \, 2}}{v_\perp^{\uparrow \, 2} + v_\perp^{\downarrow \, 2}}, \label{eq:restimation}
\end{equation}
where the upward and downward fluxes  have been expressed as proportional to the square of their respective velocity amplitudes. In turn, $\mathcal{N}$ could be computed from Equation~(\ref{eq:equifactor}) as,
\begin{eqnarray}
       \mathcal{N} &=& \frac{1}{2} + {\rm sign}\left(Z^{\uparrow}Z^{\downarrow}\right) \frac{\sqrt{F^\uparrow  F^\downarrow}}{F^\uparrow + F^\downarrow} \nonumber \\
       &=& \frac{1}{2} + {\rm sign}\left(Z^{\uparrow}Z^{\downarrow}\right) \frac{v_\perp^\uparrow v_\perp^\downarrow }{v_\perp^{\uparrow \, 2} + v_\perp^{\downarrow \, 2}}, \label{eq:nestimation}
\end{eqnarray}
 which  depends on the unknown sign of the cross term $Z^{\uparrow}Z^{\downarrow}$. This introduces an ambiguity in the estimation of the non-equipartition factor.
 
 As an example, let us consider the Hinode/SOT observations of transverse waves in spicules by \citet{Liu2014}. According to these authors, the velocity amplitudes of both upward and downward waves were determined with a method based on the two-dimensional Fourier Transform applied to height-time diagrams of the transverse displacement. The transverse velocity amplitudes reported in Table~1 of \citet{Liu2014} for cases 1 and 2 can be used to estimate the values of $\mathcal{R}$ and $\mathcal{N}$. The velocity amplitudes are $v_\perp^\uparrow \approx 2.8$~km~s$^{-1}$ and $v_\perp^\downarrow \approx 2.3$~km~s$^{-1}$ for case 1, and $v_\perp^\uparrow \approx 4.8$~km~s$^{-1}$ and $v_\perp^\downarrow \approx 3.5$~km~s$^{-1}$ for case 2. Accordingly, Equation~(\ref{eq:restimation}) yields $\mathcal{R} \approx 0.19$ for case 1 and $\mathcal{R} \approx 0.3$ for case 2. Concerning $\mathcal{N}$, the factor involving the velocity amplitudes in Equation~(\ref{eq:nestimation}) yields,
 \begin{displaymath}
     \frac{v_\perp^\uparrow v_\perp^\downarrow }{v_\perp^{\uparrow \, 2} + v_\perp^{\downarrow \, 2}} \approx 0.49,
 \end{displaymath}
 for case 1, which gives $\mathcal{N} \approx 0.01$ or $\mathcal{N} \approx 0.99$ as the two possible values of the non-equipartition factor. Similarly, for case 2,
  \begin{displaymath}
     \frac{v_\perp^\uparrow v_\perp^\downarrow }{v_\perp^{\uparrow \, 2} + v_\perp^{\downarrow \, 2}} \approx 0.476,
 \end{displaymath}
so that   $\mathcal{N} \approx 0.024$ or $\mathcal{N} \approx 0.976$. The frequency of the observed waves in these cases is approximately $10$~mHz. Comparing with the numerical results of Section~\ref{sec:atmosphere}, the estimated $\mathcal{R}$  align reasonably well with the values shown in Figure~\ref{fig:factorscontour}(a) in the upper chromosphere at similar frequencies, while Figure~\ref{fig:factorscontour}(b) suggests that $\mathcal{N} \approx 1$ at those heights. Among the possible values of $\mathcal{N}$ estimated from the observations, the largest ones agree most closely with our numerical results. This suggests that the net energy in the observed waves is predominantly kinetic in nature. Then, the net energy flux can be computed by using these estimated values of $\mathcal{R}$ and $\mathcal{N}$ in Equation~(\ref{eq:netenergy2}), along with a measure of the  phase speed of the waves, which can be associated to $v_{\rm A}$, and adopting a reasonable mass density value.

In the case that the velocity amplitudes of upward and downward waves are not available or cannot be disentangled from the observations, it is crucial to exercise caution when applying Equation~(\ref{eq:netenergy2}) due to the substantial variation of both $\mathcal{R}$ and $\mathcal{N}$ in the chromosphere, as indicated by our numerical results.  A reliable estimation of the energy flux  proves to be nearly impractical in the lower chromosphere due to the oscillatory behavior of $\mathcal{N}$. Conversely, above the critical height where kinetic energy becomes dominant, it becomes reasonable to approximate $\mathcal{N} \approx 1$. Then, at upper heights Equation~(\ref{eq:netenergy2}) might be simplified to,
\begin{equation}
    F \approx \frac{\mathcal{R}}{2}\,v_{\rm A}\, \rho\, v_\perp^2,  \label{eq:fluxenergyapp}
\end{equation}
which still depends on $\mathcal{R}$. From Equation~(\ref{eq:fluxenergyapp}), the net energy flux of the waves can be significantly smaller than the energy flux that would be computed using the classic Equation~(\ref{eq:flux3}), depending on the specific value of $\mathcal{R}$.  At upper heights in the chromosphere the reflection factor can be as small as $\mathcal{R} \sim 10^{-3}$ for low frequencies, according to Figure~\ref{fig:factorscontour}(a). Given  the uncertainty in $\mathcal{R}$,  Equation~(\ref{eq:fluxenergyapp}) should be used to estimate a possible range for the energy flux rather than providing a definitive value. Additionally, observations might also provide the phase speed of the waves, which can be substituted for $v_{\rm A}$ in Equation~(\ref{eq:fluxenergyapp}). However, it is important to note that the observed phase speed does not correspond to the velocity at which net energy propagates, $\mathcal{R}\,v_{\rm A}$, meaning that $\mathcal{R}$ cannot be estimated from the phase speed.

Finally, it is worth noting that at coronal altitudes  reflection is much less important and energy equipartition is nearly recovered in the whole frequency range, as shown in Figure~\ref{fig:factorscontour}. Hence $\mathcal{N} \approx 1/2$ and $\mathcal{R}\approx 1$, and  the classic Equation~(\ref{eq:flux3}) offers a reasonable approximation to estimate the energy flux of coronal waves.


\section{Concluding remarks} \label{sec:conclusion}

The process of continuous reflection has significant implications for the energy content of Alfv\'en waves propagating through the stratified lower atmosphere. Equipartition of kinetic and magnetic energies, a characteristic of unidirectional Alfv\'en waves in a uniform plasma, is not maintained when partial reflection occurs in a stratified medium. Among other effects, reflection reduces the velocity at which net energy propagates to values lower than the local Alfv\'en speed.  

We have considered only the effects of density stratification while neglecting magnetic field expansion. However, the main conclusions presented here about the role of reflection would remain essentially unchanged if field-line expansion was included. The combined influence of density stratification and field-line expansion determines the steepness of the Alfv\'en speed profile in the lower atmosphere and, consequently, the degree of wave reflection. The presence of field-aligned flows, which are not discussed here, may also play a role.

The analysis focuses on longitudinal stratification and does not account for transverse structuring of the magnetic field. In the lower atmosphere, the magnetic field is likely organized in flux tubes, where Alfv\'enic waves differ from classic or bulk Alfv\'en waves \citep[see, e.g.,][]{goossens2012,solerbook2024}. In this context, the effect of transverse structuring on the Alfv\'enic energy flux needs to be considered, as discussed by \citet{goossens2013} and \citet{vandoorsselaere2014}. Here, we have demonstrated that longitudinal stratification has an important impact, potentially even more pronounced than the effects of transverse structuring. Nevertheless, both longitudinal stratification and transverse structuring should be considered if a reliable estimation of the  energy content of Alfv\'enic waves in the lower solar atmosphere is attempted.

In addition to reflection, other mechanisms can contribute to the energy non-equipartition, such as Hall's effect and multi-fluid interactions \citep[see][and references therein]{martinez2025}. While these mechanisms may play a significant role in the solar wind or at higher frequencies than those analyzed here, they are not expected to be relevant for the observed waves in the lower atmosphere.

\begin{acknowledgments}
 This publication is part of the R+D+i project PID2023-147708NB-I00, funded by MCIN/AEI/10.13039/501100011033 and by FEDER, EU. The author acknowledges Marcel Goossens, Richard Morton, and Tom Van Doorsselaere for reading the manuscript and giving useful comments, and David Mart\'inez-G\'omez for providing some references.
\end{acknowledgments}




\bibliography{refs}{}
\bibliographystyle{aasjournal}

\end{document}